\documentclass[10pt,prb,aps,twocolumn,showpacs]{revtex4}
\usepackage{graphicx}

\usepackage{amssymb}

\begin{document}

\title{Variational ground states of 2D antiferromagnets in the valence bond basis}

\author{Jie Lou}
\affiliation{Department of Physics, Boston University, 
590 Commonwealth Avenue, Boston, Massachusetts 02215}

\author{Anders W. Sandvik}
\affiliation{Department of Physics, Boston University, 
590 Commonwealth Avenue, Boston, Massachusetts 02215}

\begin{abstract}
We study a variational wave function for the ground state of the two-dimensional 
$S=1/2$ Heisenberg antiferromagnet in the valence bond basis. The expansion 
coefficients are products of amplitudes $h(x,y)$ for valence bonds connecting spins
separated by $(x,y)$ lattice spacings. In contrast to previous studies, in which a
functional form for $h(x,y)$ was assumed, we here optimize all the amplitudes for 
lattices with up to $32 \times 32$ spins. We use two different schemes for optimizing 
the amplitudes; a Newton/conjugate-gradient method and a stochastic method which requires
only the signs of the first derivatives of the energy. The latter method performs significantly 
better. The energy for large systems deviates by only $\approx 0.06\%$ from its exact value 
(calculated using unbiased quantum Monte Carlo simulations). The spin correlations are also 
well reproduced, falling $\approx 2\%$ below the exact ones at long distances. The amplitudes 
$h(r)$ for valence bonds of long length $r$ decay as $ r^{-3}$. We also discuss some 
results for small frustrated lattices.
\end{abstract}

\date{\today}


\maketitle

\section{Introduction}

The valence bond (VB) basis for singlet states of quantum spin systems was first discussed by 
Rumer \cite{rum32} and Pauling \cite{pau33} in the 1930s and was shortly thereafter applied in 
Heisenberg spin chain calculations by Hulth\'en.\cite{hul38} For a systems of $N$ $S=1/2$ 
spins the overcomplete and non-orthogonal VB basis consists of states that are products of 
$N/2$  spin pairs forming singlets. In the most general case the members of a singlet can be 
separated by an arbitrary distance, but it is often convenient to consider a restricted set
of configurations with only bonds connecting two different groups of sites, e.g., the two sublattices 
of a bipartite lattice. Such a restricted VB basis is still overcomplete and any singlet state can be 
expanded in it. Any restriction on the maximum length of the bonds will render the basis incomplete, 
however. 

After Anderson's proposal in 1987 of a resonating-valence-bond (RVB) state \cite{fez74}
as a natural starting point for understanding high-temperature superconductivity 
in the cuprates,\cite{and87} variational VB states were investigated for both doped and 
undoped antiferromagnets.\cite{lia88,gros,capriotti,bon91,fano,havilio1,havilio2,poilblanc} 
The RVB spin liquid mechanism is based on states dominated by short valence bonds, which in the 
extreme case have been argued to correspond closely to the quantum dimer model.\cite{kiv87} It was 
early on established, however, that the ground state of the two-dimensional (2D) Heisenberg model 
with nearest-neighbor interactions, which has a N\'eel ordered ground state and describes 
very well the undoped cuprates,\cite{chn,man91} actually requires an  algebraic, not exponential, 
decay of the bond-length probability.\cite{lia88}

Some attempts were made to use the VB basis as a framework for numerical calculations, 
\cite{isk87,koh88,tan88,sut88,lia90,fano} but, with very few exceptions,\cite{san99} these 
efforts were not pursued further in large scale quantum Monte Carlo calculations (QMC). However, 
it was recently pointed out that there are previously unnoticed advantages of carrying out 
ground state projector QMC calculations in the VB basis, including a natural way to access excitations 
in the triplet sector.\cite{san05,beachvbs,beachinprep} Such a scheme has already been applied to 
2D and 3D models with valence bond solid ground states.\cite{vb2d,vb3d}

It may also be worthwhile to pursue further variational schemes in the VB basis,
especially exploring possibilities to study frustrated systems this way. In this paper we 
report a bench-mark variational calculation going beyond previous VB variational studies \cite{lia88} 
of the 2D Heisenberg model. Instead of assuming a functional form for the bond-length amplitudes 
and optimizing a few parameters, we optimize all individual amplitudes in order to definitely 
establish the properties of this kind of wave function and its ability to reproduce the
ground state of the 2D Heisenberg model. We also report some preliminary studies of a frustrated system.

For the standard 2D Heisenberg model with nearest-neighbor coupling $J$,
the energy of our best optimized wave function deviates by only $\Delta E/J \approx 0.06\%$ from the 
exact ground state energy for system with up to $32 \times 32$ spins. The size dependence of 
$\Delta E$ shows that this accuracy should persist in the thermodynamic limit. The error
is only half that of the best wave functions found in the previous variational QMC study.\cite{lia88} 
The spin-spin correlations are also remarkably well reproduced; they are approximately $2\%$ smaller 
than the exact values at long distances, corresponding to an $\approx 1\%$ underestimation of the
sublattice magnetization. We find that the asymptotic form of the amplitudes for bonds of length 
$r$ is $h(r) \sim r^{-3}$, which has also been found recently in a mean-field calculation.
\cite{beachmeanfield} We also compare the variational wave function with the exact ground state 
in the case of a  $4\times 4$ lattice, and find that the overlap is $\approx 0.9998$. Extending the $4\times 4$ 
calculation to a frustrated system, the J$_{\rm 1}$--J$_{\rm 2}$ model, we find that the quality of 
the amplitude-product wave function deteriorates as the frustration is increased but the overlap
remains above $0.996$ even for $J_2/J_1$ as high as $0.4$.

In Sec.~II we define the variational VB wave function.
Technical details of the QMC based optimization methods that we have used to minimize the energy 
are presented in Sec.~III. We discuss both a standard Newton method and a stochastic scheme,
which requires only the first derivatives of the energy. The latter method performs significantly 
better for large lattices. Results for the energy and spin correlations in the standard non-frustrated 
Heisenberg model are discussed in Sec.~IV. Results for energies and overlaps for the $4\times 4$ 
frustrated lattice are discussed in Sec.~V. In Sec.~VI we conclude with a brief summary and 
discussion of the methods and results.

\section{Model and wave function}

We study the standard $S=1/2$ Heisenberg model;
\begin{equation}
H = J\sum_{\langle i,j\rangle} {\bf S}_i \cdot {\bf S}_j,
\label{ham}
\end{equation}
where $\langle i,j\rangle$ denotes nearest-neighbor sites 
on a 2D square lattice and $J > 0$. The basic properties of this model have been known for a long 
time,\cite{man91} and ground state parameters such as the sublattice magnetization, the
energy, and the spin stiffness have been extracted to high precision in many QMC
studies.\cite{lia90,qmc2d,san97}  Here our aim is to investigate how well a simple variational 
wave function can reproduce the true ground state.

The general form of a VB wave function for $N$ $S=1/2$ spins is
\begin{eqnarray} 
| \Psi \rangle & = & \sum_{k}f_k | (a_1^k,b_1^k)
\cdots(a_{N/2}^k,b_{N/2}^k)\rangle \cr 
& = & \sum_{k} f_k \mid V_k \rangle ,
\label{psigeneral}
\end{eqnarray}
where \((a_i^k,b_i^k)\) represents a singlet formed by the spins at sites $a$ and $b$
in VB configuration $k$;
\begin{equation}
(a,b)=\frac{1}{\sqrt{2}}(\uparrow_a \downarrow_b - \downarrow_a\uparrow_b).
\label{singlet}
\end{equation}
The notation $| V_k \rangle $ has been introduced in (\ref{psigeneral}) for 
convenience. In the
most general case, the VB configurations $V_k$ include all the possible pairings
of the $N$ spins into $N/2$ valence bonds. A more restricted, but still massively
overcomplete basis is obtained by first dividing the sites into two groups, $A$ 
and $B$, which in the case of a bipartite lattice naturally correspond to the two 
sublattices. Here we will use such a restricted VB basis and always (also when considering 
the frustrated case later on) take $A$ and $B$ to refer to the sublattices of the square 
lattice. We fix the ``direction" of the singlet in (\ref{singlet}) by always taking the 
first index in $(a,b)$ from $A$ and the second one from $B$. With this convention one can 
show that all the expansion coefficients $f_k$ [where $k=1,\ldots,(N/2)!$] in Eq.~(\ref{psigeneral}) 
can be taken positive. This corresponds to the Marshall sign rule for a non-frustrated system
in the basis of eigenstates of the $S^z_i$ operators.\cite{lia88,sut88} 

We consider expansion coefficients of the amplitude-product form previously introduced
and studied by Liang, Doucot and Anderson;\cite{lia88}
\begin{equation}
f_k=\prod_{i=1}^{N/2} \tilde h(a_i^k,b_i^k) = \prod_{i=1}^{N/2} h(x_i^k,y_i^k),
\label{hproduct}
\end{equation}
where $x_i$ and $y_i$ are the $x$ and $y$ separations (number of lattice constants) 
between sites $a_i,b_i$, which are connected by valence bond $i$. Considering the lattice 
symmetries (we use periodic boundary conditions), there are hence $\approx N/16$ 
independent amplitudes $h(x,y)$ to optimize. In the previous study,\cite{lia88} 
only a few short-length amplitudes were optimized and beyond these an asymptotic 
form depending only on the length $r$ of the bond was assumed.\cite{manhattan}  
With a power-law form, $h(x,y) \sim r^{-p}$, it was found that long-range Neel 
order requires $p < 5$. The best variational energy was obtained with $p=4$, giving a 
deviation $\Delta E/J \approx 0.0008$ (or $\approx 0.12\%$) from the exact ground 
state energy (the thermodynamic-limit value of which is \cite{san97} 
$\approx 0.69944$ per site).

We here optimize all \( h(x,y) \) using two different methods: A standard Newton method 
(combined with a conjugate gradient method---the Fletcher Reeves method\cite{var}), and a 
stochastic method that we have developed which requires only the signs of the first derivatives. 

\section{Optimization methods}

We now discuss the technical details of these calculations. To optimize the energy
using a Monte Carlo based scheme, we write its expectation value as
\begin{equation}
E = \langle \Psi | H | \Psi \rangle = 
\frac{\sum_{kl} f_kf_l \langle V_k | V_l\rangle \frac{\langle V_k | H| V_l 
\rangle}{\langle V_k | V_l \rangle }}{\sum_{kl} f_kf_l \langle V_k | V_l \rangle},
\label{energyexp}
\end{equation}
where on the right-hand side we have taken into account the fact that we do not
normalize the wave-function coefficients, i.e., the amplitudes in Eq.~(\ref{hproduct}) 
are only determined up to an over-all factor. The overlap $\langle V_k | V_l \rangle$ between two VB 
states is related to the loops forming when the two bond configurations are superimposed; 
$\langle V_k| V_l \rangle = 2^{N_l-N/2}$, where $N_l$ is the number of loops.\cite{lia88,sut88} 
Matrix elements of ${\bf S}_i \cdot {\bf S}_j$ are also easy to evaluate in terms of the loop 
structure. If $i$ and $j$ belong to the same loop, 
then $\langle V_k | {\bf S}_i \cdot {\bf S}_j | V_l \rangle/ \langle V_k | V_l \rangle$
$=\pm 3/4$ ($+$ for $i,j$ on the same sublattice and $-$ else), and else the matrix element 
vanishes.\cite{lia88,sut88} Recently more complicated matrix elements have also been related
to the loop structure.\cite{beachvbs} 

For a given set of amplitudes $h(x,y)$, we evaluate the energy using the Metropolis Monte Carlo 
algorithm described in Ref.~\onlinecite{lia88}. An elementary update of the bond configuration 
amounts to choosing two next-nearest-neighbor sites $a,c$ (or in principle any $a,c$ in the same 
sublattice, but the acceptance rate decreases with increasing distance between the sites), 
and reconfiguring the bonds $(a,b),(c,d)$ to which they are connected, according 
to $(a,b)(c,d) \to (a,d)(c,b)$ (where the order of the labels here correspond to both 
sites $a$ and $c$ being in sublattice A). The Metropolis acceptance probability $P$ for such an update
is very easy to calculate in terms of amplitude ratios and the change in the number of
loops, $\Delta N_l$, in the overlap graph;
\begin{equation}
P = {\rm min}\left 
[\frac{h(x_{ad},y_{ad})h(x_{cb},y_{cb})}{h(x_{ab},y_{ab})h(x_{cd},y_{cd})}2^{\Delta N_l},1
\right ].
\end{equation}

\subsection{Newton Conjugate Gradient Method}

For the optimization we also need derivatives of the energy with respect to the amplitudes.
The Newton method requires first and second derivatives. Moving in a certain direction ${\bf \hat g}$
in amplitude space, the amplitude vector ${\bf h_n}$ is updated from iteration $n$ to $n+1$ 
according to
\begin{equation}
{\bf h}_{n+1}={\bf h}_n-\frac{E_g'({\bf h}_n)}{E_g''({\bf h}_n)}{\bf \hat{g}},
\label{iter2}
\end{equation}
where $E'_g$ and $E''_g$ are the first and second derivatives of the energy 
along the ${\bf \hat g}$ direction. They are calculated from 
the derivatives with respect to the amplitudes $h(x,y)$, which are evaluated during 
the sampling of VB configurations. Writing the energy expectation value as
\begin{equation}
\langle E \rangle = \frac{ \sum_{p} W_pE_p}{\sum_p W_p},~~~
W_p=\prod_{x,y} h(x,y)^{n_{xy}},
\end{equation}
where $n_{xy}$ is the total number of VBs of size $(x,y)$ in the VB configurations 
$V_k$ and $V_l$, the Monte Carlo estimator for the first derivative is
\begin{equation}
\frac{\partial \langle E \rangle}{\partial h_a}=
\left \langle\frac{n_a}{h_a}E\right\rangle-
\left\langle\frac{n_a}{h_a}\right\rangle \left\langle E \right\rangle .
\end{equation}
Here, to simplify the notation, we use $a$ as a collective index for $(x,y)$. The
second derivatives---the elements of the Hessian matrix---are:
\begin{eqnarray} 
\frac{\partial^2 \langle E \rangle}{\partial h_a^2}&=&\frac{1}{h_a^2}
\Bigl ( \langle n_a^2 E\rangle-\langle n_a^2\rangle\langle E \rangle - 
\langle n_a E\rangle+\langle n_a\rangle\langle E \rangle \nonumber \\
& & +2\langle n_a\rangle^2\langle E \rangle-
2\langle n_a\rangle\langle n_a E\rangle \Bigr ), \label{hessian1} \\
\frac{\partial^2 \langle E \rangle}{\partial h_ah_b}
&=&\frac{1}{h_ah_b}\Bigl (\langle n_an_b E\rangle-
\langle n_an_b\rangle\langle E \rangle+
2\langle n_a\rangle\langle n_b\rangle\langle E \rangle \nonumber \\
& & - \langle n_a\rangle\langle n_b E\rangle - 
\langle n_b \rangle\langle n_a E\rangle \Bigr ).
\label{hessian2}
\end{eqnarray}
Since we have many amplitudes $h(x,y)$ to optimize, we choose our optimization 
direction by the conjugate gradient method. The first direction is the 
gradient direction. In subsequent steps we choose the direction conjugate to the 
former one, satisfying the relation
\begin{equation}
{\bf h}_{n+1}\cdot {\bf A} \cdot {\bf h}_n={\bf 0},
\end{equation}
where ${\bf A}$ is the Hessian matrix. In practice, we have found that the number of line 
optimizations required for energy convergence is of the same order as total number of 
different amplitudes. Since the optimization is based on quantities obtained using a stochastic 
scheme, the final results of course have statistical errors. 

We have $\propto N^2$ different second derivatives and hence computing all of them requires 
a significant computational effort. Their statistical fluctuations are also relatively large 
for large lattices. We have also used a minimization method requiring only the first derivatives,
where the amplitudes are iterated according to
\begin{equation}
{\bf h}_{n+1}=\frac{E'_g({\bf h}_{n}){\bf h}_{n-1}-E_g'({\bf h}_{n-1}){\bf h}_{n}}
{E'_g({\bf h}_{n})-E'_g({\bf h}_{n-1})}.
\label{iter1}
\end{equation}
This amplitude update, which gives the exact minimum in the quadratic regime, works well when 
$h(x,y)$ are getting close to their optimum values. However, we still need the second derivatives
until we get very close to the optimum and in practice the advantage of using (\ref{iter1})
instead of (\ref{iter2}) at the final stages does not appear to be significant. 

We here note that a method recently proposed to reduce the statistical fluctuations of
the Hessian in variational QMC optimizations of electronic wave functions \cite{umrigar} is not 
applicable here. The proposal was to symmetrize the Hessian and write it solely in terms of 
covariances, by adding terms which are zero on average but reduce the statistical fluctuations 
of a finite sample. However, in our case the Hessian (\ref{hessian1},\ref{hessian2}) is already 
of this form and there is nothing more to do in this regard.

\subsection{Stochastic method}

We have developed a completely different optimization method which turns out to work much better
than the Newton method. It is a stochastic scheme which only requires the signs of the first 
derivatives. We update each amplitude $h_a$ according to 
\begin{equation}
\ln(h_a^{n+1})=\ln(h_a^n)-R\beta\cdot {\rm sign}
\left ( \frac{\partial \langle E\rangle}{\partial h_a} \right )_n,
\label{hshift}
\end{equation}
where $R$ is a random number in the range $[0,1)$ and ${\rm sign}(x) = 1$ if $x \ge 0$ 
and $-1$ for $x <0$. The parameter $\beta$ is gradually reduced, so that the amplitude
changes become gradually smaller. If this ``annealing" is performed slowly enough the scheme 
converges the system to the lowest energy. In fact, it turns out that the random factor $R$ 
is not even needed; the scheme converges even if $R=1$. With $R=1$, this kind of updating scheme 
has in fact been introduced previously in the context of neutral networks and is known as 
{\it Manhattan learning}.\cite{peterson,leen} Note that even in this case there is still in 
principle some randomness in the method, because when the derivatives become small the signs 
of their QMC estimates can be wrong due to statistical fluctuations. Thus, there are occasional 
adjustments of amplitudes in the wrong direction. We have found that the random factor $R$ 
speeds up the convergence, especially in the initial stages of optimization, and so we normally 
include it. 

Our method is also related to what is known as {\it stochastic optimization},\cite{robbins,stochchapter} 
where the parameter vector is updated in a steepest-decent fashion according to the stochastically
evaluated gradient;
\begin{equation}
\ln(h_a^{n+1})=\ln(h_a^n)-\beta \left ( \frac{\partial \langle E\rangle}{\partial h_a} \right )_n.
\label{gradshift}
\end{equation}
This method has been used by Harju {\it et al.}\cite{harju} to optimize electrinic wave functions. We 
have tested it on the present problem but find that it performs significantly worse than our own
variant of stochastic optimization. The reason appears to be that the fluctuations in the gradient can 
be very large, which can cause large detremental jumps in the configuration space. In our scheme the
step size is bounded and we avoid such problems.

As in simulated annealing methods, a slow enough reduction of $\beta$ should give the optimum 
solution. In the present case, unlike in simulated annealing, the optimum reached should only be 
expected to be a local optimum, although the stochastic nature of the scheme does allow for some 
more extensive exploration of the parameter space than with deterministic schemes. In stochastic 
optimization using the gradient, it has been argued that an annealing scheme of the form
\begin{equation}
\beta_k = \frac{1}{k^\alpha},~~ \frac{1}{2} < \alpha < 1,
\end{equation}
should be used in order for the method to converge. We have found this scheme with $\alpha \approx 3/4$ 
to work well.

It would be interesting to see whether this very simple scheme could also be applied to 
optimize wave functions in electronic structure calculations---the time savings from not having
to calculate second derivatives are potentially very significant in problems with a large
number of parameters.

\section{results}

\begin{figure}
\includegraphics[width=8.4cm,clip]{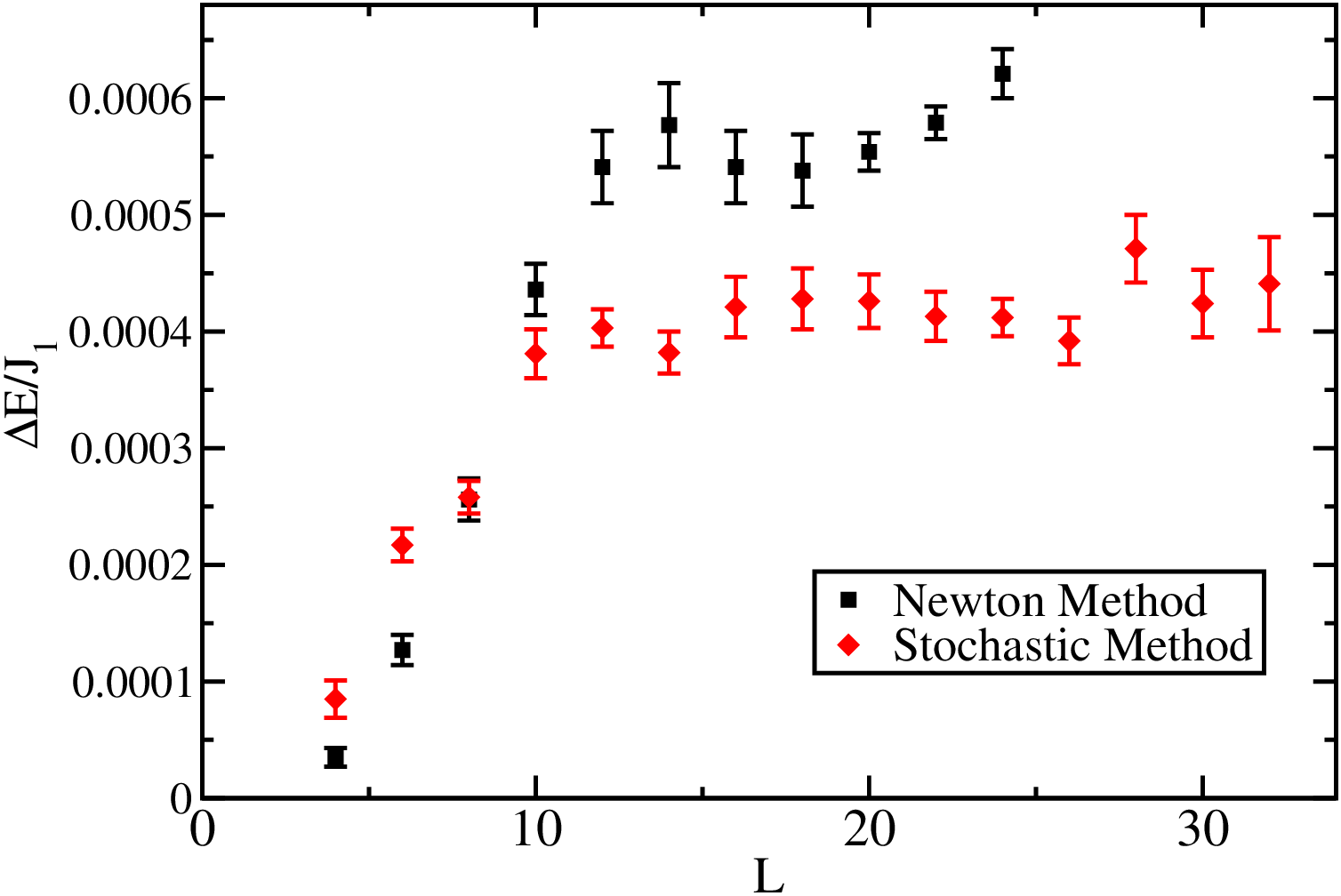}
\caption{(Color online) The deviation $\Delta E = E_{\rm var}(L)-E_{\rm exact}(L)$ of the variational 
ground state energy from the exact energy as a function of lattice size. Results obtained with
both the Newton method and the stochastic opitimization scheme are shown. The error bars
were obtained by carrying out several independent optimization runs for each $L$.}
\label{ED}
\end{figure}

We now discuss our results. In Fig.~\ref{ED} we compare the ground state energy as calculated
with the stochastic and Newton methods. We show the deviations from the correct ground
state energies (obtained using unbiased QMC calculations \cite{san97}) as a function of the lattice
size $L$ up to $L=32$. Both optimization methods give very small energy deviations for the 
$4\times 4$ system---about $0.005\%$---but the error grows as the lattice size increases.
We have calculated error bars by repeating the optimizations (from scratch) several times. 
It should be noted, however, that the fluctuations in this kind of nonlinear problem do not 
necessary have expectation value zero. Hence error bars calculated in the standard way in general 
only reflect partially the actual errors. 

For $L > 10$ the stochastic 
method delivers significantly lower energies, indicating that the Newton method has difficulties in 
locating the optimum exactly. We believe that this problem is due to the statistical errors of the 
second derivatives (which are much larger than those of the first derivatives). Convergence issues 
related to statistical fluctuations of the Hessian are well known in variational calculations for 
electronic systems.\cite{umrigar} Any systematic shifts in the Newton method would of course be reduced 
by increasing the length of the simulation segments used to calculate the energy and its derivatives
in each iteration. Simulations sufficiently long to reach the same level of optimization as with
the stochastic method do not appear to be practically feasible, however. In the stochastic scheme all 
statistical
errors should decrease to zero as the cooling rate is reduced. We cannot, of course, guarantee that the 
results shown in Fig.~\ref{ED} are completely optimal, but we have carried out 
the simulations at different cooling rates in order to check the convergence. Based on these tests we do 
believe that the results are converged to their optimum values to within the error bars shown. 

For the larger lattices, the energy deviation in Fig.~\ref{ED} is only $\Delta E/J\approx 0.0004$, 
or $\approx 0.06\%$, and is size independent within statistical errors for $L >10$. This should then be 
the accuracy in the thermodynamic limit. In Ref.~\onlinecite{lia88} only a few of the short-length amplitudes 
were optimized and a functional form---power-law or exponential---was used for the long-range behavior. 
The best power-law  wave functions had energy deviations of $\Delta E/J \approx 0.0008$; twice as large 
as we have obtained here with the fully optimized amplitudes.

\begin{figure}
\includegraphics[width=7.9cm,clip]{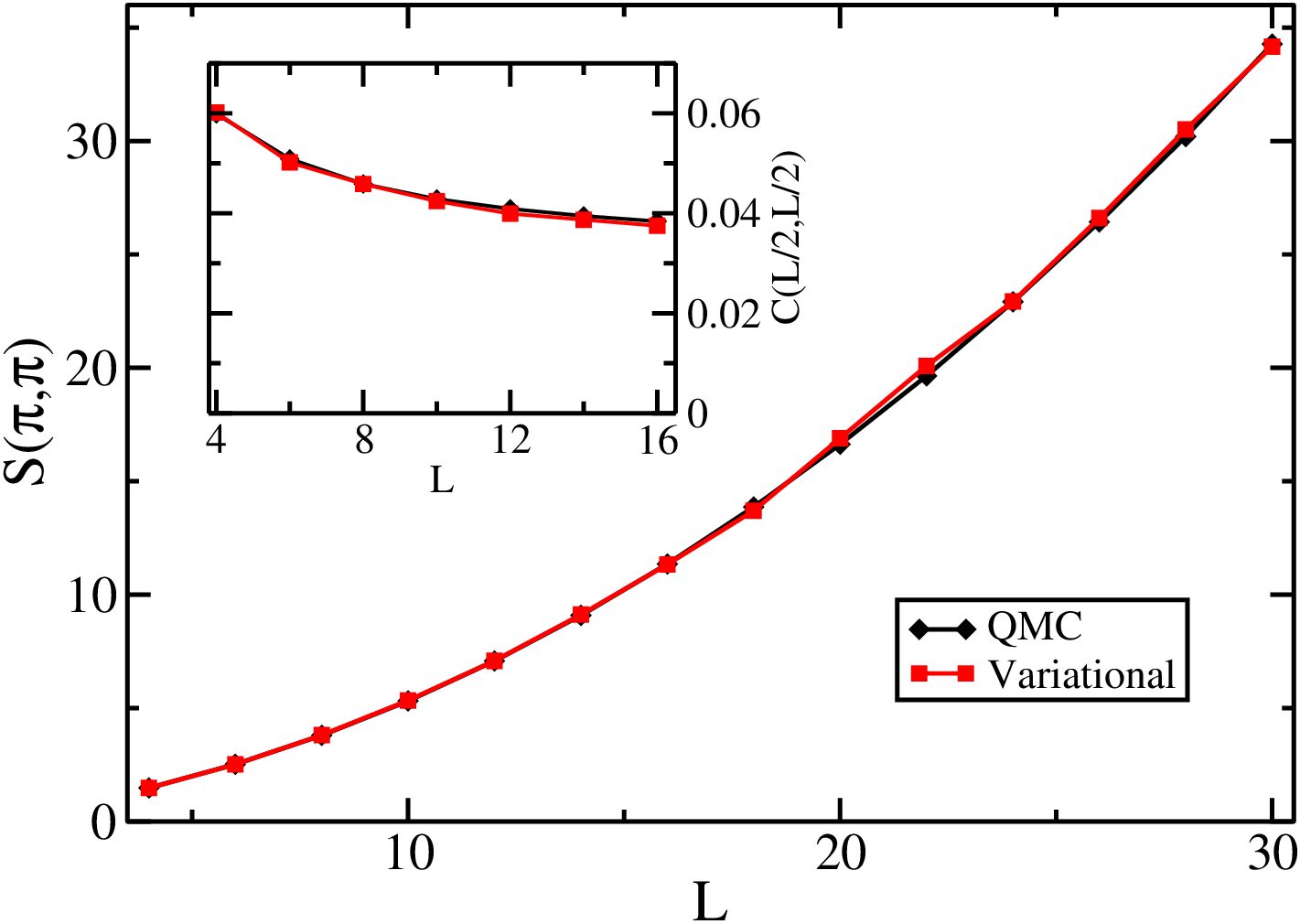}
\caption{(Color online) The staggered structure factor $S(\pi,\pi)$ versus lattice size,
compared with unbiased QMC results. The inset shows the long-distance spin-spin correlation. Statistical
errors are smaller than the symbols.}
\label{SC}
\end{figure}

Having concluded that the stochastic method is the preferred optimization technique, we discuss 
only results for other quantities obtained this way. Fig.~\ref{SC} shows the size dependence
of the staggered structure
factor,
\begin{equation}
S(\pi,\pi) = \sum_{x,y} (-1)^{x+y} C(x,y),
\end{equation}
where $C(x,y)$ is the correlation function, defined by
\begin{equation}
C(x_i-x_j,y_i-y_j)= \langle {\bf S}_i \cdot {\bf S}_j\rangle.
\end{equation}
The inset shows the correlation function at the longest distance; $(x,y)=(L/2,L/2)$,.
We again compare with results from unbiased QMC calculations.\cite{san97}
The structure factor of the variational ground state agrees very well with the exact 
result for these lattice sizes---the deviations are typically less than $0.5\%$. The long-distance 
correlations show deviations that increase slightly with $L$, going to $\approx 2\%$ below the
true values for $L \ge 10$ [which then should also be the asymptotic $L \to \infty$ error of
$S(\pi,\pi)$]. The sublattice magnetization is the square-root of the long-distance 
correlation function; it is thus only $1\%$ smaller than the exact value.

\begin{figure}
\includegraphics[width=7cm,clip]{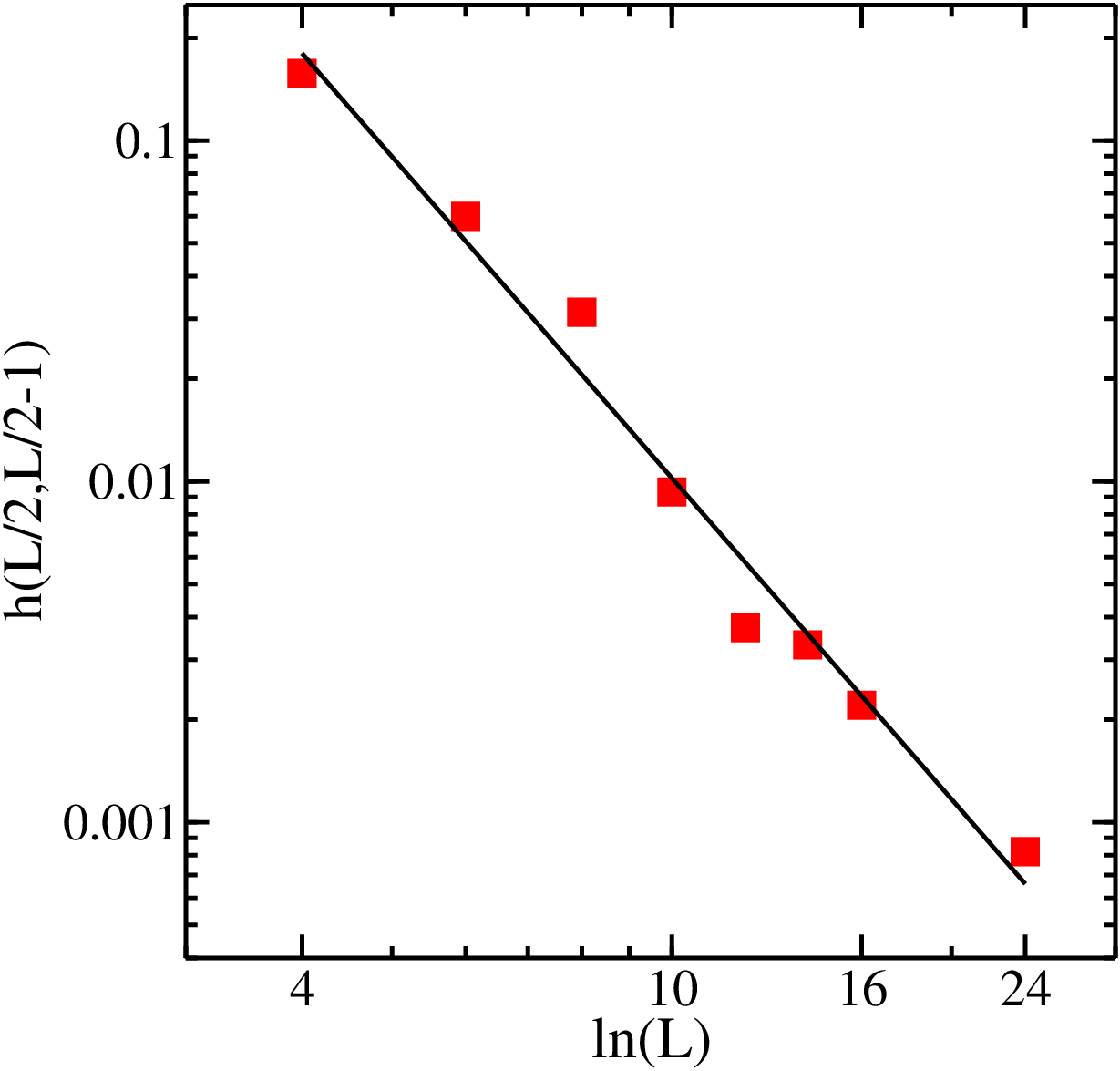}
\caption{Log-log plot of the amplitude $h(L/2,L/2-1)$ versus the system size. 
Statistical errors are of the order of the size of the circles. The line 
shows the power-law $h \sim L^{-3}$.} 
\label{HR}
\end{figure}

Liang et al.~did not conclusively settle the question of the asymptotic behavior of the 
amplitudes $h(x,y)$ for bonds of long length $r=(x^2+y^2)^{1/2}$.\cite{lia88}
The best variational energy was obtained with an algebraic decay; $h \sim r^{-4}$. 
However, the energy is not very sensitive to the long-distance behavior and the values 
obtained for $r^{-3}$ and $r^{-2}$ were not substantially different at the level of statistical 
accuracy achieved. Even with an exponential decay of the bond-length distribution the energy 
was not appreciably higher, but then no long-range order is possible and hence this form
can be excluded. In a recent unbiased projector QMC calculation, the probability distribution 
$P(x,y)$ of the bonds was calculated.\cite{san05} The form 
$P(r) \sim r^{-3}$ was found (with no discernible angular dependence). Without a hard-core 
constraint for the VB dimers, the probabilities would clearly be proportional to the amplitudes; 
$P(x,y) \propto h(x,y)$, and even with the hard-core constraint one would expect the two to be 
strongly related to each other. In fact, as was pointed out in Ref.~\onlinecite{san05}, a wave 
function with $h(r) \sim r^{-p}$ does result in $P(r) \sim r^{-p}$. Our variational calculation 
confirms that indeed the fully optimized $h(r) \sim r^{-3}$, as demonstrated in Fig.~\ref{HR} 
using the longest bonds, $(x,y)=(L/2,L/2-1)$, on the periodic lattices. 

Havilio and Auerbach carried out a VB mean-filed calculation which gave an
exponent $p \approx 2.7$.\cite{havilio2} The statistical accuracy in Fig.~\ref{HR} is 
perhaps not sufficient to definitely conclude that $p=3$ exactly, or to exclude $p=2.7$, from 
this data alone. However, the QMC study of the probability distribution $P(R)$ supports $p=3$ 
to significantly higher precision.\cite{san05}  Moreover, Beach has recently developed a 
different mean-field theory which predicts $p=d+1$ for a $d$-dimensional system.
\cite{beachmeanfield} There is thus reason to believe that $r^{-3}$ indeed is the 
correct form for $d=2$.

\begin{figure}
\null\hskip2mm
\includegraphics[width=8.4cm, clip]{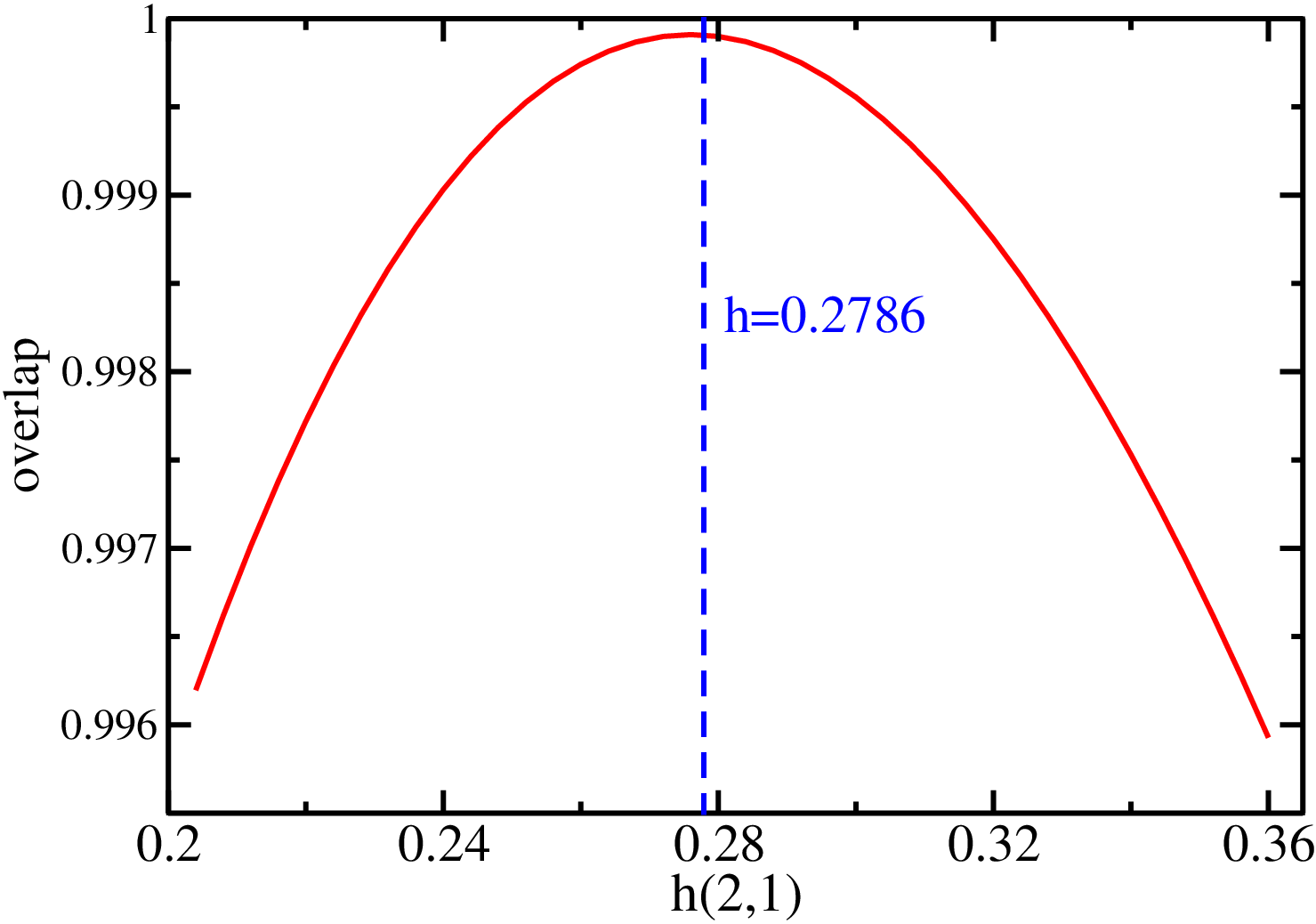}
\caption{(Color online) Overlap between the exact $4\times 4$ wave function and the 
VB wave function versus the single independent amplitude $h(2,1)$. The value of $h(2,1)$ 
obtained in the variational calculation is indicated.}
\label{Overlap}
\end{figure}

For the $4\times 4$ lattice we can compare the variational wave function with the
exact ground state obtained by exact diagonalization. This comparison is most easily 
done by transforming the VB state to the $S^z$ basis. Taking into account 
the lattice symmetries, there are $822$ $m_z=0$ states with momentum $k=0$, and
the matrix can easily be diagonalized. We generate the $8!$ VB states $| V_k \rangle$ 
using a permutation scheme and convert each of them into $2^8$ $S^z$-basis states with 
weights $\pm \prod h(x,y)$, and use these to calculate the overlap with the exact
ground state. With the amplitudes normalized by $h(1,0)=1$, there is only one 
independent amplitude, $h(2,1)$, to vary for the $4\times 4$ lattice. In 
Fig.~\ref{Overlap} we show the overlap as a function of $h(2,1)$. We also 
indicate the value of $h(2,1)$ obtained in the variational QMC calculation---it 
matches almost perfectly that of the maximum overlap. The best overlap is indeed 
very high; $\approx 0.9998$. It would be interesting to see how the overlap
depends on the system size. For a $6\times 6$ lattice, the ground state can also be 
calculated, using the Lanczos method, but the space of valence bond states is too 
large to calculate the overlap exactly (although it could in principle be done by 
stochastic sampling).

\section{Frustrated systems}

We have also studied the Heisenberg hamiltonian including a frustrating interaction;\cite{j1j2}

\begin{equation}
H = J_1\sum_{\langle i,j\rangle} {\bf S}_i \cdot {\bf S}_j +
J_2\sum_{\langle\langle i,j\rangle\rangle} {\bf S}_i \cdot {\bf S}_j,
\label{fham}
\end{equation}
where $\langle i,j\rangle$ and $\langle\langle i,j\rangle\rangle$ denotes nearest and next-nearest
neighbors, respectively, and $J_1,J_2 > 0$. Also in this case there exists, in principle, a 
positive-definite expansion of the ground state in the valence bond basis. This can be easily seen
because a negative coefficient $f_k$ in Eq.~(\ref{psigeneral}) can be made positive simply by 
reversing the order of the indices in one singlet in that particular state. However, no practically 
useful convention for fixing the order is known. We here use the same partition of the lattice into 
A and B sublattice sites as in the non-frustrated case and the same sign convention (\ref{singlet}) for the 
singlets. We only consider the $4\times 4$ lattice, which, as we will show, already gives some 
interesting information on the behavior of the simple amplitude-product wave function as the 
frustration ratio $J_2/J_1$ is increased.

\begin{figure}
\includegraphics[width=8.4cm, clip]{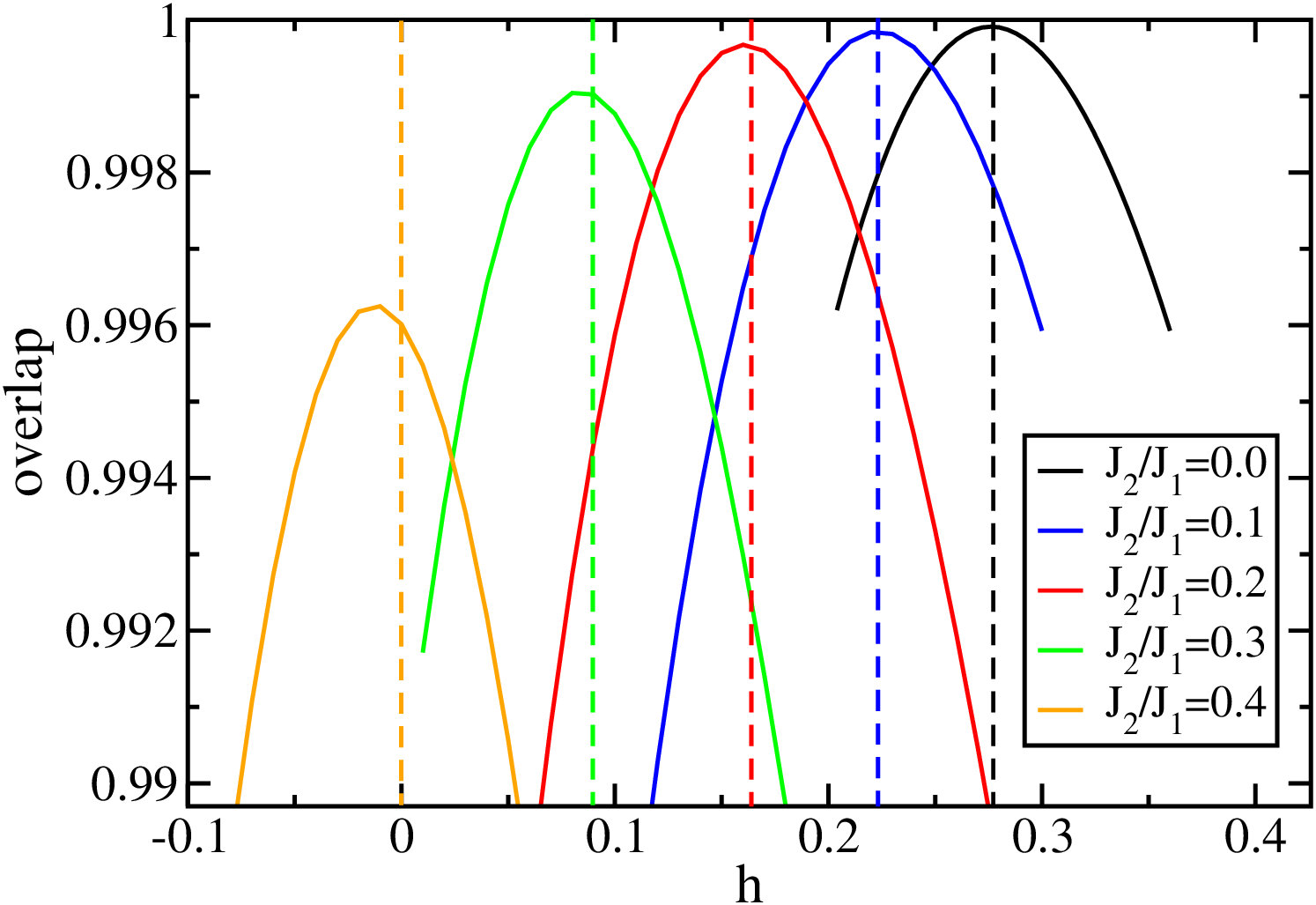}
\caption{(Color online) Overlap beween the exact $4\times 4$ ground state and the VB wave 
function for different values of the frustration $J_2/J_1$. The best amplitudes obtained in
variational Monte Carlo calculations are indicated with the dashed lines.}
\label{Overlap2}
\end{figure}

In the exact calculation we can study both positive and negative values of $h(2,1)$, but for now 
we restrict the variational calculation to $h(2,1) > 0$, in order to avoid the Monte Carlo sign 
problem caused by negative amplitudes.\cite{signnote} It should be noted, however, that the sign problem here
is much less severe than in exact QMC schemes,\cite{san05} and hence there is some hope of actually 
being able to consider mixed signs in variational QMC calculations in the VB basis.\cite{signnote}

In Fig.~\ref{Overlap2} we plot the dependence on $h(2,1)$ of the overlap between the  VB wave 
function and the exact ground state for several values of $J_2/J_1$. The $h(2,1)$ corresponding 
to maximum overlap decreases as the frustration increases. For $J_2/J_1=0.4$ the best overlap 
occurs for $h(2,1)<0$. The optimum overlap decreases significantly with $h(2,1)$ for $J_2/J_1 \agt 0.3$, 
indicating the increasing effects of bond correlations not taken into account in the product-form 
of the expansion coefficients. This deterioration of the wave function may be related to the
phase transition taking place in this model at $J_2/J_1 \approx 0.4$.\cite{j1j2} Note, however,
that even at $J_2/J_1=0.4$ the overlap remains as high as $\approx 0.996$.

There is a point close to $J_2/J_1=0.4$ where $h(2,1)$ vanishes and thus the best wave function
for the $4\times 4$ lattice contains only bonds of length $1$. 
Beyond this coupling the optimum wave function requires a negative
$h(2,1)$. It has also been noted previously that wave functions including only the shortest 
bonds give the best description of the ground state in a narrow region of high frustration 
in a model containing also a third-nearest neighbor interaction $J_3$.\cite{mambrini}

We also show in Fig.~\ref{Overlap2} the values of $h(2,1)$ obtained in the variational calculations. 
Interestingly, these values coincides well with the maximum overlaps only when the frustration is weak, 
showing that the best variational state in a given class is not always the best in terms of the wave 
function.

\section{Summary and conclusions}

In conclusion, we have shown that a variational valence bond wave function, parametrized in terms 
of bond amplitude products, gives a very good description over-all of the 2D Heisenberg model.
Although this has been known qualitatively for a long time,\cite{lia88} our study shows that the agreement
is quantitatively even better than what was anticipated in previous studies. The deviation of the 
ground state energy from the exact value is $\approx 0.06\%$ for large lattices; almost $50\%$ better 
than in previous calculations where a functional form was assumed for the amplitudes.\cite{lia88} 
The sublattice magnetization is correct to within $\approx 1\%$ (smaller than the true value).
We have also shown that the amplitudes for bonds of length $r$ decay as $r^{-3}$ for large $r$, which 
is the same form as for the probability distribution calculated previously.\cite{san05} It is also 
in agreement with a recently developed mean-field theory.\cite{beachmeanfield}

By exactly diagonalizing the hamiltonian on a $4\times 4$ lattice, we have also studied the 
frustrated J$_{\rm 1}$--J$_{\rm 2}$ model. Not surprisingly, we found that the quality of the 
amplitude-product wave function deteriorates when the frustration 
$J_2/J_1$ is increased. However, even at $J_2/J_1=0.4$, i.e., 
close to the phase transition taking place in this model,\cite{j1j2} the
overlap is above $0.996$. It would clearly be interesting to see how well the amplitude-product
state works for the frustrated model on larger lattices. In this regard we note that Capriotti 
{\it et al.} \cite{capriotti} recently carried out a variational study of an RVB function written 
in terms of fermion operators \cite{and87} and found that it gave the best description of the ground state 
of the J$_{\rm 1}$--J$_{\rm 2}$ model at large frustration; $J_2/J_1 \approx 0.5$. However, the
overlap is significantly smaller than what we have found here for the $4\times 4$ system at
the same level of frustration.

Although the fermionic \cite{and87} and bosonic descriptions of the VB states are formally equivalent, 
the fermionic wave function, as it is normally written, does not span the full space possible with
the bosonic product state. As a consequence, the bosonic description we have used here 
can in practice deliver much better variational wave functions for non-frustrated systems.\cite{poilblanc} 
The fermionic description apparently works better for frustrated than non-frustrated systems.\cite{capriotti} 
However, if the sign is also optimized for each amplitude in the bosonic product state (which is not 
easy for large highly frustrated systems, however, because of Monte Carlo sign problems \cite{signnote}) 
it is clear that these wave functions should be better than the fermionic RVB state considered so 
far.\cite{capriotti} Therefore, the VB wave function we have studied here should, at least in 
principle, give an even better description of the ground state of the frustrated model than the 
fermionic RVB state optimized in Ref.~\onlinecite{capriotti}. Our results for the $4\times 4$ lattice, 
along with the results of Ref.~\onlinecite{capriotti}, suggest that the QMC sign problem \cite{signnote}
should be small up to $J_2/J_1 \approx 0.4$. It may thus even be possible to gain insights into the quantum 
phase transition and the controversial state\cite{j1j2} for $J_2/J_1 > 0.4$ with this type of 
variational wave function.

Another interesting question is how bond correlations, which are not included in the wave function 
considered here, develop as the phase transition at $J_2/J_1 \approx 0.4$ is approached. We are currently 
exploring the inclusion of bond-pair correlations to further improve the variational wave function for 
the Heisenberg model as well as more complicated spin models. 

The stochastic energy minimization scheme that we have introduced here, which requires only the signs 
of the first energy derivatives, may also find applications in variational QMC simulations of 
electronic systems. Recently proposed efficient optimization schemes \cite{umrigar,sorella} need 
the second energy derivatives, and so our scheme requiring only first derivatives has the potential 
of significant time savings when the number of variational parameters is large. Very recently, other
powerful schemes also requiring only the first energy derivatives have been developed and have been 
shown to be applicable to wave functions with a large number of parameters.\cite{umrigar2} 
We have not yet compared the efficiencies of these different optimization approaches with the 
stochastic scheme presented here.

\acknowledgments

We would like to thank Kevin Beach for many useful discussions. This work was supported by the 
NSF under Grant No.~DMR-0513930.


\begin{thebibliography}{00}
\bibitem{manhattan}
The ``Manhattan distance" $r=|x|+|y|$ was used, but very similar results are obtained 
with $r=(x^2+y^2)^{1/2}$.
\bibitem{signnote}
In the fermionic representation of the VB states \cite{and87} there is no sign problem in 
variational calculations \cite{poilblanc,capriotti} because the sampling is done using orthogonal 
basis states. For the VB states we use here, the non-orthogonality leads to sign problems when 
frustrated interactions are present.
\end{thebibliography}
\end{document}